\documentclass[twocolumn,showpacs,floatfix,superscriptaddress]{revtex4}
\usepackage{latexsym}
\usepackage{graphicx}
\usepackage{dcolumn}
\usepackage{bm}
\usepackage{latexsym}
\usepackage{booktabs}
\usepackage{amsmath}
\usepackage{amssymb}
\usepackage{float}
\usepackage{setspace}
\usepackage[english]{babel}
\usepackage{hyperref}
\usepackage{amsfonts}
\usepackage{psfrag}
\usepackage{babel}
\usepackage{array}
\newcommand{\bq}{\begin{equation}}
\newcommand{\eq}{\end{equation}}
\newcommand{\ba}{\begin{eqnarray}}
\newcommand{\ea}{\end{eqnarray}}
\usepackage{color}

\begin{document}

\title{On the fate of dynamical systems under a trade-off between cost and precision}
\author{Maximilian Voit}
\affiliation{Physics and Earth Sciences, Jacobs University Bremen, P.O.Box 750561, 28725 Bremen, Germany.}
\author{Hildegard Meyer-Ortmanns}\email[]{h.ortmanns@jacobs-university.de}
\affiliation{Physics and Earth Sciences, Jacobs University Bremen, P.O.Box 750561, 28725 Bremen, Germany.}

\begin{abstract}
We analyze the fate of dynamical systems that consist of two kind of processes. The first type is supposed to perform a certain function by processing information at a required high accuracy, which is, however, limited to less than 100 percent, while the second process serves to maintain the required precision. Both processes are assumed to be subject to a trade-off between cost and precision, where the cost have to be paid from renewable but limited resources. In  a discrete map we pursue the time evolution of errors and determine the conditions under which the fate of the system is either a stable performance at the desired accuracy, or a deterioration. Deterioration may be realized either as an accumulation of errors or a decline of resources when they are all absorbed for maintenance. We point to possible implications for living organisms and their perspectives to avoid an accumulation of errors in the course of time.
 \end{abstract}


\maketitle

In recent years it turned out to be  constructive  to consider biological processes from the viewpoint of information theory. Biological systems like genetic networks, signaling proteins and cells sense, transduce and process internal signals and external signals from the environment to optimize their behavior, to use the limited resources for their own benefit \cite{tkacik}. These processes may be summarized under information-processing dynamics and biological computation. Thus it is natural to assess biological processes in terms of accuracy, energetic efficiency and speed of performance.

Information transfer is realized in the transcription and translation machinery, in molecular motors or pumps and enzymatic reactions. For these biomolecular processes a thermodynamic uncertainty relation was formulated in \cite{seifert1} and proven in \cite{ginrich}. It states that a more precise output (such as the number of product molecules generated by an enzymatic reaction) requires a higher thermodynamic cost (independent of the time used to produce the output). The product of the total dissipation after time $t$ and the square of the relative uncertainty of the measured observable satisfy an inequality relation so that the product is always larger or equal to $2k_B T$ with $T$ the temperature and $k_B$ the Boltzmann constant. 

Furthermore, from a comparison between cellular copy protocols with canonical copy protocols in computational science it became clear \cite{tenwolde1,tenwolde2,tenwolde3} why cellular sensing systems can never reach the Landauer limit \cite{landauer,bennett} on the optimal trade-off between accuracy and energetic cost.
Cellular systems have to dissipate more than thermodynamic processes that run according to ideal quasistatic protocols. 
Cellular copy protocols can only reach $100\%$ precision for diverging costs.

Admittedly, different sources of stochastic fluctuations need not automatically
add up to a noisy output of reactions. In particular, negative feedback can both amplify and suppress
noise \cite{brugge}; however, noise in one
part of the feedback loop is reduced at the expense of an increased level
of noise in another part of the loop.
Moreover, correlations play an important role in noise propagation
through the different layers of a network, from the genetic to the cellular
level. Anti-correlations between extrinsic noise from the environment and intrinsic noise inside the cells can improve the robustness of biochemical networks \cite{tenwoldecorr}. Apart from such physical mechanisms, nature has invented a repertoire
of sophisticated repair mechanisms like kinetic proofreading \cite{hopfield} to counteract a rapid accumulation of errors. Theoretically the proofreading can achieve infinite specificity the longer the reaction runs, but there at the cost of large amounts of the (correct) product as well. So also proofreading seems to be subject to a trade-off between cost and efficiency.

Based on these results we extrapolate toward the conjecture that information-processing biological systems are universally subject to trade-offs between cost and precision, though the precise form of the cost-precision function may depend on the specific process. Thus we expect them  to perform at less than 100 percent precision in general. Otherwise
we abstract from concrete realizations and propose a simple model to pursue the time evolution of error fractions under such an assumed inherent trade-off.

The model is supposed to project on basic mechanisms in a more general class of dynamical systems, which are composed of two type of processes.
In common to processes of the first class is their requested high-precision performance, since they otherwise lose their function. This may be a function  on the microscopic level in the biological context, or related to fabrication and production processes on a macroscopic level.
As representative for the first class we choose a simple cyclically repeating copying process of a bitstring of length N that is initially correct with all bits set to zero. When the defective copies iteratively get further copied, errors accumulate. Each bit is copied with precision $p_0$  that equals the probability of copying a bit correctly. The parameter $p_0$ denotes the initial value of maximal precision, typically chosen as $p_0=0.90$.\\
The second class of processes serves for maintenance of the first class, here represented as repair processes of erroneously copied bits, once the error fraction $x$ exceeds a certain tolerable error threshold $\zeta$. The repair is successful with probability $\rho\le 1$, so $\rho=1$ corresponds to deterministic repair, $\rho< 1$  to probabilistic repair. Thus, on average, $\rho$ gives the fraction of bits for which repair was attempted and successful.  We keep $\rho$  either constant, or let it decrease with time in steps of size $d$. \\
Both kinds of processes need resources, for simplicity we assume only one type of resource, termed ``energy". For the energy costs per copy of a single bit at highest precision $p_0$ we choose $c(p_0)=f/N >1$, where $f$ denotes the costs of copying the whole string. The actual value of $f$ depends on the choice of the cost-precision function $c(p)$ (that will be specified below) and on the value for the initial precision. The repair costs of a single bit are chosen as $R\cdot c(p_0)$ with $R\ge 0$.

The assumed fundamental trade-off between cost and precision is not explicitly modeled, but effectively implemented via a cost-precision function $c(p)$ for copying a single bit with precision $p$.  The choice is only constrained by the boundary conditions such that
$\lim_{p\rightarrow 1}c(p)=\infty $ and $\lim_{p\rightarrow 0}c(p)=1$.
 The limit $p\rightarrow 1$ corresponds to perfect copying without errors and requires an infinite amount of energy, while random copying without any guarantee of correctness causes minimal nonzero costs set to $1$.
We test different options for $c(p)$, in particular power-like behavior
$c(p)=1/(1-p^{n_1})^{n_2}$
for $n_1=1$, $n_2=1$ or $2$, or $n_1=2$, $n_2=1$ as well as a logarithmic dependence according to
$c(p)=1-\log(1-p)$.\\
 We equip the system with a temporary energy reservoir per ``sweep" that can be refilled: When we allow for correcting erroneously copied bits beyond the threshold $\zeta$, one sweep amounts to the repair of a string with a fraction $x$ of errors exceeding the tolerated threshold $\zeta$ and the subsequent copying of all bits. As capacity of the energy reservoir we choose
$f=N\cdot c(p_0)$,
so that initially enough energy is available to copy a bitstring just once and with highest available precision $p_0$. We shall keep the capacity of size $f$ constant over time. This means that for each sweep (representing a fixed finite time interval), always the same finite amount of energy is accessible, independently of how the reservoir was depleted in the preceding sweep and independently of its actual large size. Even if its size would be practically infinite, in real systems it is always a finite amount that can be extracted in a finite time. Our minimal choice as $f=Nc(p_0)$  is for simplicity; adding initially a constant reserve would delay its depletion, letting it fluctuate may delay or accelerate its depletion.

Importantly, the energy needed for the correction of the bitstring is taken from the very same energy reservoir that also serves to provide energy for the copying process. Since we assume a direct correlation between copying precision and energetic costs, the precision decreases as soon as energy has to be spent on repair. We  analyze the conditions under which  a stationary performance of the copying process is still possible that maintains the desired accuracy  below the tolerated error threshold $\zeta$. Alternative fates of the system are complete deterioration in the sense that either errors accumulate in the bitstring, so that the error fraction $x$ approaches 1, or the copying process terminates before it is completed. This happens when the energy supply is insufficient and all energy is spent on repair, before the next copying process starts.

The previous steps can be summarized in the following algorithm which stops when all available energy is used up by repair, and the system is said to be in a state of deterioration. The highest initial precision $p_0$, the tolerated error fraction $\zeta$, the costs-precision function $c(p)$, the cost factor $R$ for repair, the probability $\rho$ for a successful repair, and the decline $d$ of the success probability $\rho$ are kept fixed for the whole simulation.\\
1. Initialization: choose $c(p_0)$, $p_0$, $\zeta$, $R$, $\rho$, $d$. Start from a bitstring $\{0,...,0\}_b$ of zeros.\\
2. Copy the string with initial precision $p_0$.\\
 3. Compute the current error fraction $x\in[0,1]$, i.e., the fraction of bits that are $1$. Also, set the energy available for this step to $N\cdot c(p_0)$.\\
 4. If $x<\zeta$ go to 2, 
 else  (for $x\ge\zeta$ and $d\ge 0$) decrease $\rho$ by $d$ and go to 5.\\
5. Make an attempt to repair $(x-\zeta)\cdot N$ bits, with success probability $\rho\le 1$. (This means that on average $\rho(x-\zeta)\cdot N$ randomly chosen bits actually get repaired.) Even if a repair is not successful, the attempt of repair causes $R\cdot c(p_0)$ energy cost per bit, so that the remaining energy for the subsequent copying process decreases to $N\cdot c(p_0)(1-(x-\zeta)R)$.\\
6.  If $N\cdot c(p_0)(1-(x-\zeta)R)\le 0$, stop, else go to 7.\\
 Use the rest of accessible energy for copying the whole string with highest possible precision $p=c^{-1}(c(p_0)(1-(x-\zeta)R))\le p_0$ (with $c^{-1}$ the inverse function) and go to 3.\\
Typical choices are $c(p)=\frac{1}{1-p}$, $p_0=0.9$, $\zeta=0.1$; $R$ and $\rho$ will be used as bifurcation parameters. Important special cases are deterministic repair, that is, $\rho=1$ and $d=0$, and constant repair success rate with $\rho<1$ and $d=0$.

\begin{figure}[ht]
  \includegraphics{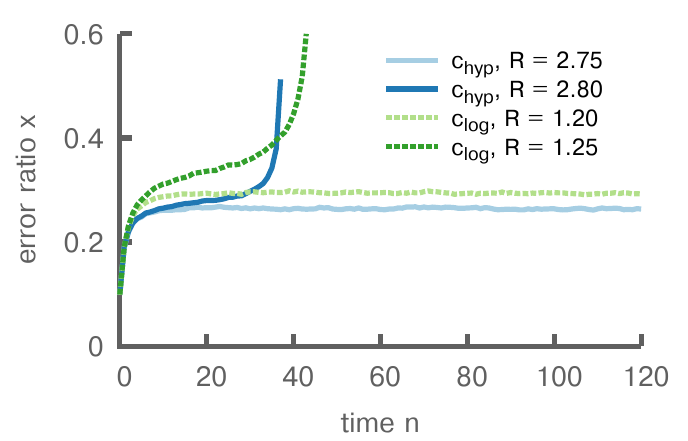}
  \caption{Numerical time evolution $x(n)$ for two different cost
    functions ($c_{hyp}(p) = 1/(1-p)$ and $c_{log}(p) = 1-ln(1-p)$),
    each one for $R < R^*$ and $R > R^*$.
    Other parameters are $p_0 = 0.9$, $\zeta = 0.1$, $N = 100000$.  }
  \label{fig8}
\end{figure}

Figure~\ref{fig8} shows the time evolution of the error fraction $x$ for two repair cost factors $R$, for $\rho=1$ and two cost-precision functions $c(p)=\frac{1}{1-p}$ and $c_{\log}(p)=1-\log(1-p)$. As a first observation, the qualitative behavior of the error fraction  sensitively depends on $R$, approaching a constant value $x_n<1$ for $R$ below some ``critical" value $R^*$, or rapidly increasing towards $1$ above this critical value. The critical parameter range, where the qualitative behavior of $x_n$ changes, depends on the choice of $c(p)$.

Along with an increasing error fraction
goes a decreasing precision $p(n)$  and an increasing fraction $\psi(n)$ of energy which is used for repair, again for values of $R$ below and above $R^*$.

In the limit of large $N$ we may use the probabilities for erroneous copies and successful repair as the actual mean values. For deterministic repair ($\rho=1$) we obtain the discrete map from $x_n$ at time step $n$ to $x_{n+1}$ at time step $n+1$
\begin{equation}
x_{n+1}=\zeta+(1-\zeta)(1-c^{-1}(c(p_0)(1-(x_n-\zeta)R)))\;,
\label{eq14}
\end{equation}
when a fraction of $\zeta$ errors are tolerated, the remaining $(1-\zeta)$ fraction of the bits may get wrong during copying with precision $p$. This generates $(1-\zeta)\cdot(1-p)$ erroneous copies, where the precision
\begin{equation}
p=c^{-1}(c(p_0)(1-(x_n-\zeta)R)),
\end{equation}
with $c^{-1}$  the inverse cost-precision function, is determined such that the energy costs per bit
$c(p)$ can be exactly covered by the remaining energy per bit after repair.
Note that the discrete map holds only for $x_{n}\ge\zeta$, which excludes the (linear) transient at early times when errors are allowed to accumulate toward the threshold that is assumed to be tolerable.

\begin{figure}[ht]
  \centering
  \includegraphics{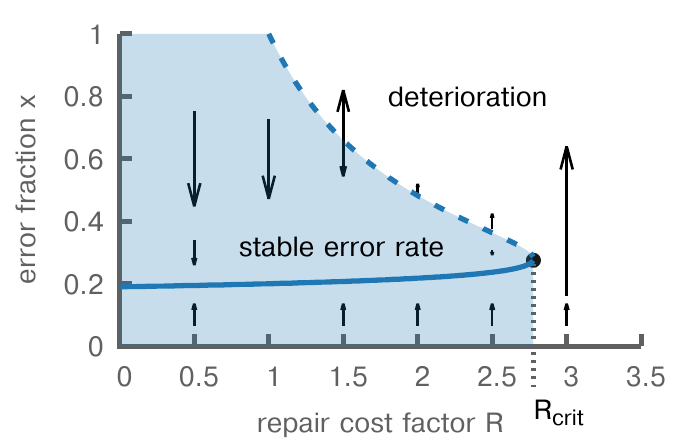}
  \caption{Bifurcation diagram for parameter values $p_0 = 0.1, \zeta = 0.9$.
    Plotted are the location of the stable (solid line) and unstable (dashed line) fixed
    points against the size of the repair-cost factor $R$.
}
  \label{fig12}
\end{figure}

For $\rho\le 1$ and the cost-precision function $c(p)=\frac{1}{1-p}$ the map becomes
\begin{eqnarray}\label{eq:map1}
x_{n+1}&=&\zeta + (x_n-\zeta)(1-\rho)\nonumber\\
&+& (1-\zeta-(x_n-\zeta)(1-\rho))\tfrac{1-p_0}{(1-(x_n-\zeta)R)}.
\end{eqnarray}
Here $(x_n-\zeta)(1-\rho)$ is the fraction of errors which were repaired without success, while from the remaining bits a fraction of $1-p=\frac{1-p_0}{(1-(x_n-\zeta)R)}$ gets erroneously copied.
\begin{widetext}
Fixed points of this map are obtained from setting $x_{n+1}=x_n$:
\begin{equation}\label{eq99}
x^*_{\pm}=\frac{1}{2R\rho}(1+p_0(\rho-1)+2R\rho\zeta)
\pm\sqrt{1+p_0^2(\rho-1)^2+2p_0(\rho-1-2R\rho(\zeta-1))+4R\rho(\zeta-1)}.
\end{equation}
\end{widetext}
Since $x \in \mathbb{R}$, these are only physically relevant under the condition that the square root is real. For $\rho=1$ this condition leads to
\begin{equation}
R\le R_{\text{crit}}=\frac{1}{4(p_0(\zeta-1+1)-\zeta)}.
\end{equation}
The stability analysis of these fixed points shows that $x_+^*$ is unstable, while $x_-^*$ is stable.
 Figure~\ref{fig12} displays the location of both fixed points as a function of the repair cost factor $R$ for $\rho=1$.
The solid (dashed) line marks the locations of the stable (unstable) fixed points, respectively. Since only values of $x<1$ are physically meaningful, there exists only a stable fixed point in the shaded area between $R=0$ and $R=1$. For $R>1$ we find a coexistence between a stable and an unstable fixed point for given $R$. States within the shaded region to the left of and below the unstable branch are stable in the sense that from all locations in this area the error fractions converge towards the line of stable fixed points. The fate of the system then is a stable repeating copying process at constant error rate given by the stable fixed point that is approached. \\
In the coexistence regime it depends on the initial value of $x$ whether the system approaches a stable fixed point or ends up in the phase of deterioration, where the error rate of the copying process rapidly approaches $1$. Even from the area below the line of unstable fixed points, stochastic fluctuations may kick the system across the line into the regime of deterioration.

A fold bifurcation happens at $R_{\text{crit}}=\frac{1}{4(p_0(\zeta-1+1)-\zeta)}$ (black dot in Figure~\ref{fig12}), where stable and unstable fixed point collide and vanish. For $R > R_{\text{crit}}$ the system always deteriorates.

\begin{figure}[ht]
  \includegraphics{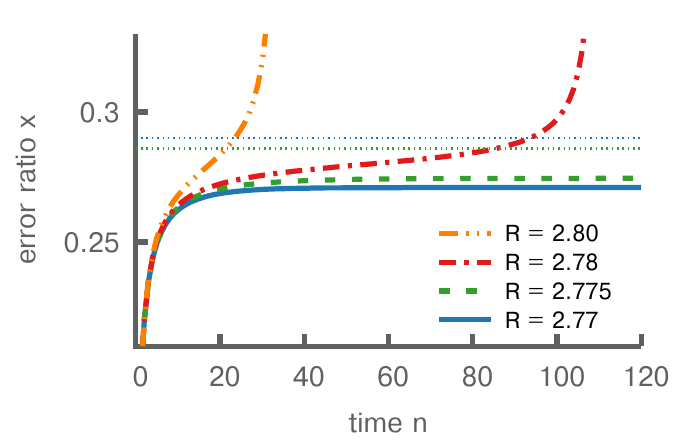}
  \caption{Mean-field error fraction $x(n)$ as a function of time for different repair costs $R$. Dotted lines mark the location of the unstable fixed points.
  Other parameters are $p_0 = 0.9$, $\zeta = 0.1$.  }
  \label{fig14}
\end{figure}

Figure~\ref{fig14} shows the time evolution of the error fraction $x$ for different repair costs factors $R$, for $\rho=1$ and $c(p)=\frac{1}{1-p}$, above and below $R_{\text{crit}}$. Similarly to the numerically determined time evolution of Figure 1, the error fraction sensitively depends on $R$, approaching a constant value $<1$ (solid curve), or slowly (red/dark-grey curve, $R=2.78$) or rapidly (orange/light-grey curve $R=2.8$) increasing towards $1$. The slow increase of errors for a small interval of $R$ is due to the ``felt" vicinity of the stable fixed point. If a stochastic fluctuation pushes the (otherwise stable) system beyond the unstable fixed points, it will deteriorate. As soon as the system leaves the vicinity of this fixed point, $x$ rapidly approaches $1$, so that the string consists of only erroneous bits. The dotted lines mark the unstable fixed points (that only exist for $R < R_{\text{crit}}$) for the two $R$-values below $R_{\text{crit}}$.
Note that the analytically calculated value of $R_{\text{crit}}=2.\bar7$ for $\rho=1$ and $c(p)=\frac{1}{1-p}$ reproduces quite well the numerically observed threshold.

Stochastic fluctuations about the mean-field results for finite $N$ can go in both directions (figure not displayed):
After a rapid initial increase of the error rate, followed by a transient stabilization if $R$ is close to $R_{\text{crit}}$ the system gets deteriorated  towards $x=1$ at a time that can be earlier or later than the mean-field prediction  with stronger fluctuations for smaller values of $N$.

\begin{figure}[ht]
  \centering
  \includegraphics{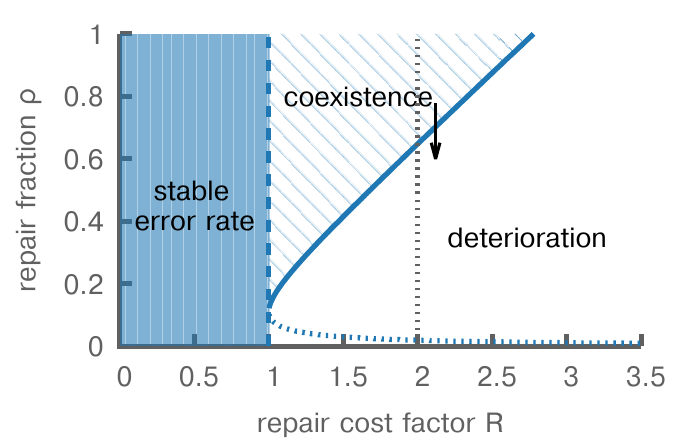}
  \caption{Bifurcation diagram  as a function of
    varying repair probability (repair success) $\rho$ and repair cost factor $R$.
    The solid line marks the critical value $R_{\text{crit}}$ of fold bifurcations, continued by the blue/darkgrey dotted line. In the shaded region the system approaches a stable error rate, below the solid line the error fraction goes to $1$. In the striped coexistence region both outcomes are possible.
    The vertical dotted line marks a possible path when the repair success $\rho$ is decaying ($d > 0$). When the system crosses the line of fold bifurcations from above, it will deteriorate. Other parameters are $p_0 = 0.1, \zeta = 0.9$.
}
  \label{fig16}
\end{figure}

For  variable $\rho\le 1$, both $\rho$ and $R$ serve as bifurcation parameter. The formerly single fold bifurcation point becomes the line with a solid and dashed branch in Figure~\ref{fig16},
analytically obtained from the vanishing of the $\pm$ square root in Eq.~\ref{eq99} with the solid (dashed) branch for $+\sqrt{}$ ($-\sqrt{}$), respectively. The solid and dashed branches meet at $R=1$, since for our choice of parameters $p_0+\zeta=1$ and $\sqrt{}=0$. For $R<1$ and $\rho>0$, all initial conditions for $x$ lead to a stable fixed point with constant error fraction. In the striped area the fate of the system depends on the initial value for $x$. The small area below the dashed line of fold bifurcations corresponds to non physical values for $x$.
Finally, the vertical dotted line marks a possible way for fixed $R$ that a system with decaying repair success would take towards the phase of deterioration.  Fig.~\ref{fig12} corresponds to a $\rho=1$-slice of this figure.

\begin{figure}[ht]
  \centering
  \includegraphics{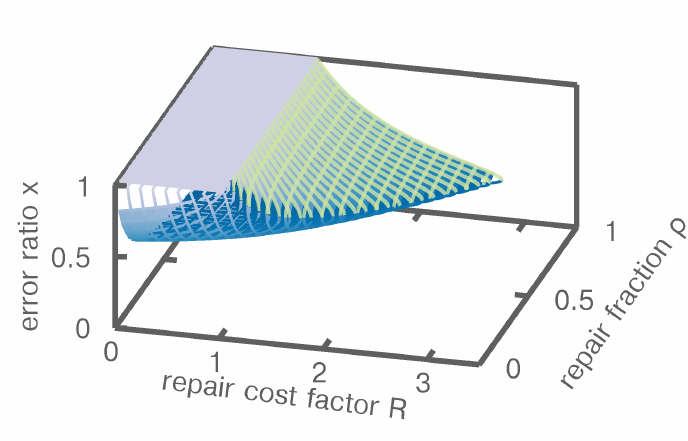}
  \caption{Bifurcation diagram in the $R$-$\rho$-$x$ space.
    Figure~\ref{fig12} is a $\rho=1$ cut of this plot.
      }
  \label{fig17}
\end{figure}

The three-dimensional plot of Figure~\ref{fig17} shows the volume between the surfaces of stable
(blue/dark-grey) and unstable (green/light-grey) fixed points.
The crest where both surfaces meet is the line of fold bifurcations. From any point below this volume the system is attracted to a stable fixed point in the error fraction, as long as $R$ is not too large, above the volume the phase is characterized by deterioration. Figure~\ref{fig12} is a cut of Figure~\ref{fig17} through the $(x,R)$-plane for $\rho=1$.
When the repair costs are smaller than the copying costs $R<1$, it is plausible that a constant error rate can be maintained. For $R>R_{crit}$ repair absorbs all energy after a few steps, so that no energy is left for copying. On the other hand, repair without success $(\rho=0)$ and $R>1$ leads unavoidably to error accumulation, while for $\rho\not=0$ the chance for error stabilization at larger repair cost factors increases with increasing $\rho$.

So far we have assumed that repair is not perfect if $0<\rho<1$, but constant over time. If a trade-off between costs and precision also applies to the repair mechanisms, these mechanisms will get less efficient  over time with an increasing number of unsuccessful repair events, unless a superordinated repair mechanism controls and corrects the repair. A self-consistent implementation would require to account for the costs caused by the repair of repair, which we neglected so far when setting $\rho=const$. Without such an iterated control mechanism it is natural to let the repair success decrease with time, here after each sweep by an amount of $d$. Apart from the region with $R<1$ this means that starting at $\rho=1$, the surface of stable fixed points will be crossed towards the deterioration regime, unavoidably and sooner for larger $R$, later for smaller $R>1$, as seen from Figure~\ref{fig17}.

The matching between analytical and numerical predictions for the bifurcation diagram according to Figure~\ref{fig12} as well as \ref{fig16} and \ref{fig17} has been confirmed for the cost-precision function $c(p)=\frac{1}{1-p}$. Via AUTO \cite{auto} we have checked that the analogous bifurcation diagrams for our other choices of cost-precision functions
lead to qualitatively very similar $(x-R)$-diagrams as in Figure~\ref{fig12}, also when plotted for different choices of the cost-repair factor $\rho$. Therefore we expect our results as summarized in Figure~\ref{fig17} not to be  sensitive to the choice of $c(p)$ as long as the boundary conditions are the same.

In summary, even if the energy for repair and copying are taken from the same reservoir so that repair is at the expense of the subsequent copying precision, errors need not automatically approach the maximal ratio 1, but may saturate at fixed points with values that are assumed to be tolerable. The final fate depends on the actual success and costs for repair in relation to copying. It should be noticed that the cost for keeping the repair success constant have not  been explicitly included in the energy balance. If repair is also subject to a trade-off between cost and precision, such a constant success rate requires investment, otherwise the success will decline in the course of time and unavoidably lead to deterioration as displayed in our bifurcation analysis.\\
Already simple extensions of our model toward a large (rather than minimal) and fluctuating energy reservoir and further iterations of repair of repair mechanisms may considerably delay the accumulation of errors if not completely prevent it.\\
When errors accumulate, essential parts of living organisms will lose their function as soon as   certain tipping points in error fractions of the steering ``codes" are exceeded. If the accumulation of errors in biological information-processing is taken as one of the basic manifestations of organisms' aging, the extended model is able to explain the large variance in lifetimes as well as the (on average) rather long time scale, on which aging emerges. This time is usually long as compared to the intrinsic time scales of the ``production" and maintenance processes. In real biological organisms, many processes of information transfer are ongoing in parallel with their specific costs and precision, so that their overall fate seems unpredictable from first principles. However, once the cost-precision functions for sufficiently simple organisms are identified, a balance along this kind of model may in particular predict that aging is an unavoidable fate of this organism. It would be unavoidable as it is based on fundamental physical principles such as the universally existing trade-offs in the face of finite accessible resources.

Acknowledgment.  Financial support from the German Research Foundation DFG (grant no. ME-1332/28-1) is gratefully acknowledged.

{}

\end{document}